\documentclass{elsart}

 \usepackage{graphicx}

\usepackage{amssymb}

\newcommand{\ud}{\mathrm{d}}

\begin{document}

\begin{frontmatter}



\title{A one-parameter family of interpolating kernels for Smoothed Particle Hydrodynamics studies}


\author{Rub\'{e}n M. Cabez\'{o}n}
\address{Dept. de F\'{i}sica i Enginyeria Nuclear, UPC. Jordi Girona, 1-3, 08034 Barcelona, Spain}
\ead{ruben.cabezon@upc.edu}

\author{Domingo Garc\'{i}a-Senz}
\address{Dept. de F\'{i}sica i Enginyeria Nuclear, UPC. Jordi Girona, 1-3, 08034 Barcelona and Institut d'Estudis Espacials de Catalunya. Gran Capit\`a 2-4, 
 Barcelona, Spain}
\ead{domingo.garcia@upc.edu}

\author{Antonio Rela\~{n}o}
\address{Dept. de F\'{i}sica i Enginyeria Nuclear, UPC. Jordi Girona, 1-3, 08034 Barcelona, Spain}
\ead{antonio.relano@upc.edu}

\begin{abstract}
A set of interpolating functions of the type $f(v)=\left\{\sin\left[\frac{\pi}{2}v\right]/\left(\frac{\pi}{2}v\right)\right\}^n$ is analyzed in the context of the smoothed-particle hydrodynamics (SPH) technique. The behaviour of these kernels for several values of the parameter $n$ has been  studied either analytically as well as numerically in connection with several tests carried out in two dimensions.  The main advantage of this kernel relies in its flexibility because for $n=3$ it is similar to the standard widely used cubic-spline, whereas for $n>3$ the interpolating function becomes more centrally condensed, being well suited to track discontinuities such as shock fronts and thermal waves.
\end{abstract}

\end{frontmatter}

\section{Introduction}
The technique called smoothed particle hydrodynamics (SPH) was introduced in 1977 \cite{l77} and \cite{gm77} to simulate the evolution of fluids and plasmas in three dimensions. It is a gridless Lagrangian method hence one has not to be worried about the definition of a mesh and its further remapping because the grid is somehow advected by the particles themselves (see \cite{m05} for a recent review of this technique). A central point of the SPH formalism is the concept of interpolating function (or kernel) through which the continuum properties of the fluid are recovered from a discrete sample of $N$~points with mass $m_i$~which move according to the hydrodynamical laws. A good interpolating kernel must satisfy a few basic requirements: it must tend to a delta function in the continuum limit and has to be a continuous function with, at least, definite first and second derivatives. From a more practical point of view it is also advisable to deal with symmetric kernels of finite range, the latter to avoid $N^2$ calculations. 
A prototype of kernel is the Gaussian kernel: 

\begin{equation}
W^\mathrm{G}(v,h) =\frac{1}{(\sqrt\pi h)^d}\exp{(-v^2)}
\label{eq:eq1}
\end{equation}

where $d$ is the spatial dimension and $v=\vert$$\mathbf{r - r'}$$\vert/h$ is the normalized distance between particles in terms of the characteristic smoothing length $h$. The precise value of $h$ sets a spatial length-scale which is usually taken as 
the local resolution of SPH. Nevertheless the Gaussian kernel has an infinite range thus it is most practical to use a similar function but with compact support. A function with the required properties is the cubic $M_4$~spline 
\cite{s46},~\cite{ml85} defined as:

\begin{equation}
M_4(v,h) =\frac{A}{h^d}
\cases
{1-\frac {3}{2}v^2+\frac {3}{4}v^3\qquad\qquad 0\le v\le 1\cr\cr
\frac{1}{4}(2-v)^3\qquad\qquad\qquad 1<v\le 2\cr\cr
0\qquad\qquad\qquad\qquad\qquad\quad v>2\cr}
\label{eq:eq2}
\end{equation}

where $A=\frac {2}{3}$, $\frac {10}{7\pi}$ and $\frac {1}{\pi}$ in one, two and three dimensions respectively. This kernel works very well and it is computationally efficient as it has been checked worldwide by many people \cite{b90} and \cite{m92} using SPH in many areas of physics and astrophysics. 

In general, hydrodynamical simulations with SPH should not significantly depend on the chosen kernel but there could be special cases in which the precise profile of the interpolating device becomes relevant. These situations appear wherever there is an abrupt change of physical variables. Trivial examples are strong shock formation and thermal discontinuities. In these cases the values of the variables around the jump region are severely damped by the interpolation procedure and the values at the peak are underestimated. In other cases the numerical noise can become high enough to blur the true value of the variables. If there are chemical or nuclear reactions, which are very sensitive to temperature, the outcome might be greatly altered or even completely wrong. Several ways to better handle discontinuities in SPH have been proposed. One of them is to enhance the artificial viscosity algorithm of the codes in order to improve the jump of the variables at the discontinuity and its thickness  \cite{m97}. Another route is to try a more clever interpolation using kernels especially devised to handle large gradients. In particular tensorial kernels with ellipsoidal geometry were proposed by \cite{ovsm98}. In normal conditions these kernels reduce to the spherically symmetric cubic-spline but, in the presence of a shock the sphere becomes an ellipsoid with its minor axis aligned with the shock direction leading to  an improvement of the resolution. However several interesting properties of the spherically symmetrical kernels are lost during their transformation into ellipsoids. In particular, the use of spherically symmetrical kernels guarantees that any function is approximated to second order in $h$, thus linear functions are exactly reproduced. In addition energy and momentum are not so well conserved when ellipsoidal kernels are used. An alternative to tensorial interpolators is to consider strongly peaked kernels with spherical symmetry. In this case, for a fixed value of the smoothing parameter $h$, there is an increase in resolution, although that enhancement is always accompanied by an increase of the numerical noise \cite{fq96}. In this paper we present a one-parametric family of spherically symmetric kernels based on harmonic-like functions ($W^{H}_n$ kernels hereafter) with compact support. Specifically we propose that the set of functions: 
 
\begin{equation}
W^{H}_n(v,h)=B_n(h)
\cases
{1 \qquad\qquad\qquad\qquad v=0\cr\cr
\left(\frac{\sin\left[\frac{\pi}{2}v\right]}{\frac{\pi}{2}v}\right)^n\qquad 0<v\le 2\cr\cr
0\qquad\qquad\qquad\qquad v>2\cr}
\label{eq:eq3}
\end{equation}

\noindent where $B_n$~is a normalization factor, have several interesting features which make them suitable to SPH studies. A change in the value of the governing parameter $n$ lead to different shapes of the interpolative function, from more extended to more centrally condensed profiles as $n$~increases. Moreover, for $n=3$~it is very similar to the well-known cubic spline given by Eq. (2). 
Thus we suggest that the use of the $W^{H}_n$~set of functions in current SPH calculations would add  more flexibility to the numerical scheme without practically introducing any inconvenient.  Although the function given by 
Eq. (3) is of great importance in signal analysis theory, where the case n=2 
corresponds to the so-called window function, as far as we know it has 
never been used 
before in connection with SPH. A somehow related interpolator $W\propto 
 (1-v^2/4)(1+\cos\pi v/2)$~was studied by \cite{fq96} although only in 1D. 
However it was a single kernel, not a continuous family as those given by Eq.(3)

Interpolators of type $W_n^H$~with a high $n$~should be compared not only 
to the $M_4$~but to higher order splines. In this respect the 
quintic $M_6$~polynomial interpolator \cite{s46} could be taken as representative.  
It has a compact support and 
has been used to model flows using 
SPH \cite{mfz97}. As the $M_6$~spline was originally devised to work within   
a radii of $3h$~we 
have renormalized it to $2h$~in order to make plausible comparisons with $W_n^H$:

\begin{equation}
M_6(v,h) = \frac{C}{h^d}
\cases
{(2-v)^5-6 (\frac{4}{3}-v)^5+15 (\frac{2}{3}-v)^5
 \qquad 0\le v\le \frac{2}{3}\cr\cr
(2-v)^5-6 (\frac{4}{3}-v)^5\qquad\qquad\frac{2}{3}<v\le \frac{4}{3} \cr\cr
(2-v)^5\qquad\qquad\qquad \frac{4}{3}<v\le 2\cr\cr
0\qquad\qquad\qquad\qquad\qquad\quad v> 2\cr}
\label{eq:eq2}
\end{equation}

\noindent 
where $C=\frac{243}{2560}, \frac{15309}{61184\pi}$~and $\frac{2187}{10240\pi}$~in one, two and 
three dimensions respectively.

In Section 2 we give the main mathematical features of the kernels defined by Eq.~(\ref{eq:eq3}), and discuss their abilities to handle steep functions. A comparative analysis of the performance of these interpolators in 
disordered discrete systems is also given in the same section. 
In Section 3 we analyze the behaviour of $W^{H}_n$ in connection to two classical tests: (1) the propagation of a thermal wave arising from a thermal discontinuity in 2D cartesian coordinates and, (2) the evolution of a blast wave reaching the Sedov phase, also calculated in 2D. Finally a brief critical discussion concerning the virtues and shortcomings of the proposed $W^{H}_n$~set of functions and  the main conclusions of our work is provided in Section 4.

\section{General properties of the $W^{H}_n$~set}

In this section we study the relevant mathematical features of the family of kernels we are proposing. Mathematical theory of interpolation tells us that the Fourier transform of $f(w)=\left\{\sin (w)/w\right\}^k$,~ where $w=\pi h/\lambda$ can be used to generate a whole family, $M_k$, of polynomial interpolators of increasing degree \cite{s46}. For $k=2$~the Fourier transform of $f(w)$~leads to the linear interpolator $M_2$~spline whereas the cubic spline is obtained for $k=4$. Conversely the $W^H_k$~set can be regarded as the inverse Fourier transform of the $M_k$~family of polynomial kernels. Such reciprocal relationship basically arises because all acceptable kernels are merely a variation of the basic Gaussian function defined by Eq.~(\ref{eq:eq1}). Thus the $W^H_n$~set of functions can also be considered good interpolating functions with the interesting peculiarity that $W^H_3$~is practically equal to the cubic spline $M_4$~not after the Fourier transform of $f(w)$~but in current cartesian space.

A plot of the  profile of these kernels for several values of the 
governing index $n$~is shown in Fig. 1. As it can be seen the resulting profile for $n=3$~is very close to that of the cubic spline, with the additional advantage that its second derivative is smoother. Interestingly, the 
profile of the quintic $M_6$~is well fitted by $n=5$~(strictly, 
$n=4.9$~provides 
a slightly better fit) whereas the Gaussian kernel is reasonably reproduced 
by choosing $n=2$. Values of $n$~above 4 lead to condensed high-peaked kernels 
which could be useful to handle abrupt spatial changes in physical variables during current hydrodynamical calculations. 

\subsection{Normalization}
Firstly we provide the normalization factor $B_n$ for $W^H_n$ in equation 3 as: 

\begin{equation}
B_n=\frac{1}{I_1h^d}=\frac{K_n}{h^d}
\label{eq:eq4}
\end{equation}

\noindent
where $d$~is the dimension of the space and, 

\begin{equation}
I_1=\int_{\infty}\left(\frac{\sin\left[\frac{\pi}{2}v\right]}{\frac{\pi}{2}v}\right)^n 
dV
\label{eq:eq5}
\end{equation}

which has to be calculated numerically. In Table 1 there is shown the value of $K_n$ for one, two and three dimensions for different values of $n$. These values can be plotted in a diagram, Fig. 2, and approximated by a polynomial function which gives $K_n$~with very good accuracy for any value of $n$ within the  range $2\le n\le 7$,  

\begin{equation}
K_n= a_5n^5+a_4n^4+a_3n^3+a_2n^2+a_1n+a_0
\label{eq:eq6}
\end{equation}

The values of the fitting coefficients $a_i$~in 1D, 2D and 3D are provided in Table 2. The relative errors introduced by the fitting formulae are 
negligible, 
lesser than $10^{-5}$. It is also safe to extrapolate Eq. (7) down to 
$n=3/2$~and 
up to $n=8$~without introducing a significative error. Note that the normalization constants $K_n$~depicted in Fig. 2 intersect around $n=7$. For $n>7$~the 1D normalization constants always lie below those defined by the  2D and 3D lines. A plausible interpretation is that in 2D and 3D calculations with a finite 
number of particles there is a practical limit of the exponent $n$. Above that value the kernel becomes too sensible to the locus of the very firsts neighbors of the particle thus loosing their isotropy features. That is, the 3D kernel behaves pretty much as an one-dimensional interpolator. Thus we have constrained the value of $n$~to be in the range $2\le n\le 6$~in all the numerical tests given in Section 3. 

\subsection{Fitting Gaussians with $W^H_n$}

Some insight about the ability of the $W^H_n$~set to reproduce steep functions can be gained through the study of simple Gaussians curves and surfaces with sharp slopes. To begin with let us consider the following one dimensional Gaussian 
density profile: 

\begin{equation}
\rho(v)=\rho_0\left(1+R\,e^{-v^2}\right); \qquad\qquad v=x/h
\label{eq:eq7}
\end{equation}

\noindent
which represents a Gaussian curve with a characteristic width just equal to the smoothing-length parameter $h$ and maximum value  $\rho_{max}=(1+R)\rho_0$. Thus the interpolators will find some difficulty to reproduce these curves because the resolution is similar to the width of the bell. We are mainly interested in three magnitudes: the maximum peak of density, the width of the curve at a half of that maximum and the maximum value of the density gradient. These Gaussians could be viewed as idealized mathematical curves mimicking the jump of physical magnitudes such as density or temperature associated to shock waves and thermal fronts. For example, taking R=3 in Eq. (\ref{eq:eq7}) leads to a density jump of a factor four, the same that for a strong shock passing through a perfect gas with $\gamma=5/3$.

From the standard SPH definitions of interpolation of a function we have (in 1D): 

\begin{eqnarray}
\left<\rho_{max}\right>&=&\int_{-2}^{2} \rho_0\left(1+R\,e^{-v^2}\right)B_n\left(\frac{\sin\left[\frac{\pi}{2}v\right]}{\frac{\pi}{2}v}\right)^n h\ud v=\nonumber\\
&=&B_nh\rho_0\int_{-2}^{2}\left(\frac{\sin\left[\frac{\pi}{2}v\right]}{\frac{\pi}{2}v}\right)^n\ud v+B_nh\rho_0 R\int_{-2}^{2}e^{-v^2}\left(\frac{\sin\left[\frac{\pi}{2}v\right]}{\frac{\pi}{2}v}\right)^n\ud v
\label{eq:eq8}
\end{eqnarray}

calling $I_1$~and $I_2$~to the integrals on the right side of 
Eq. (\ref{eq:eq8}) we have:

\begin{equation}
\left<\rho_{max}\right>=\rho_0\left(1+R\frac{I_2}{I_1}\right)
\label{eq:eq9}
\end{equation}

Therefore the closer $I_2/I_1$~is to 1 the better the numerical approximation is. In Table 1  there is shown the fraction $I_2/I_1$~as a function of the parameter $n$~which characterizes the harmonic-like kernels for 1, 2 and 3 dimensions. As it can be seen as $n$~increases the interpolation improves substantially. For the particular case $n=3$ the density peak does not reach the $50\%$~of the true value in the 3D case whereas for $n=6$~the percent rises to $64\%$. 

Similarly, the maximum value of the derivative of Eq. (\ref{eq:eq7}) is: 

\begin{equation}
\frac{\partial{\rho}}{\partial{x}}\bigg\vert_{max}=\frac{R}{h}\rho_0\sqrt{\frac{2}{e}}
\label{eq:eq10}
\end{equation}

which takes place at $v=x/h=-\sqrt{2}/2$. The estimation of that derivative using the kernel is (see \cite{m05}~for details on how to calculate derivatives with SPH): 

\begin{eqnarray}
\left<\frac{\partial\rho}{\partial x}\right>_{x=-\sqrt{2}/2}&=&n\rho_0 B_n\int_{-2}^{2} \left(1+R\,e^{-(v+\sqrt{2}/2)^2}\right)\left(\frac{\sin\left[\frac{\pi}{2}v\right]}{\frac{\pi}{2}v}\right)^n\times \nonumber\\ 
&\times&\left[\frac{\frac{\pi}{2}}{\tan\left[\frac{\pi}{2}v\right]}-\frac{1}{v}\right]\ud v 
\label{eq:eq11}
\end{eqnarray}

It is easy to show that Eq. (\ref{eq:eq11}) reduces to:

\begin{eqnarray}
\left<\frac{\partial\rho}{\partial x}\right>_{x=-\sqrt{2}/2}=\frac{R}{h}\rho_0 \left(n\frac{I_3}{I_1}\right)
\label{eq:eq12}
\end{eqnarray}

where, 

\begin{eqnarray}
I_3=\int_{-2}^{2} e^{-(v+\sqrt{2}/2)^2}\left(\frac{\sin\left[\frac{\pi}{2}v\right]}{\frac{\pi}{2}v}\right)^n\left[\frac{\frac{\pi}{2}}{\tan\left[\frac{\pi}{2}v\right]}-\frac{1}{v}\right]\ud v
\label{eq:eq13}
\end{eqnarray}

The factor $(n I_3/I_1)$~for the 1D case as a function of $n$~is given in the last column of Table 1. As we can see approaching the derivative through kernel estimation using Eq. (\ref{eq:eq12}) leads to maximum values which are around the $55\%$~(n=3) and $71\%$~(n=6) of the analytical estimation given by Eq. (\ref{eq:eq10}). 

Finally, in Fig. 3 there is shown the complete smoothed density profile of the Gaussian and its spatial derivative for the cases $n=3, 5, 6~ \mathrm {and}~ n=8$. As we can see there is a clear enhancement of the resolution as $n$~rises, in both the function and its derivative. However the increase in resolution 
is not monotonic because the gap between  $n=3$~and $n=6$~is larger than that 
from $n=5$~to $n=8$.
We can see that, in consonance to the results above, the width of the bell at half-height also improves as the index of the kernel rises. Those cases with a 
higher index $n$~make also a better approach to the derivative of the Gaussian profile in all points, as it can also be seen in Fig. 3. It is worth noting that the ability of the  $W^H_n$~family to handle the above sharp Gaussians is {\sl independent} of the particular value adopted for $h$ (see for instance Eqs. 10, 11 and 13). Such good behaviour is due to the fact that both, the function and the kernel share the same characteristic width, $h$. In other cases the outcome of the interpolation will be dependent of the smoothing-length parameter. However such dependence is weak, of second order in $h$, \cite{m05}, thus the discussion given in this section still holds unless the width of the Gaussian becomes much lesser than the smoothing-length $h$.         

\subsection{Pairing-instability}

Pairing instability is a well-known problem of SPH \cite{sha95}. Particles that are too much close tend to clump together forming a stable configuration which is, in fact, a numerical artifact. This situation is caused by an unstable stress-strain relation that occurs typically when the second derivative of the kernel becomes negative and high strain is developed between particles. It could be of practical interest to have a direct control on the locus 
where the pairing-instability comes out. The higher the index 
$n$~in $W^H_n$~is, 
 the closer is that locus to the origin, making harder for a particle to get 
stuck with its closest neighbor.

\begin{equation}
(W^H_n)^{''}=n\left(\frac{\pi}{2}\right)^2W^H_n\left[n\left(\frac{1}{\tan\left(\frac{\pi}{2}v\right)}-\frac{1}{\frac{\pi}{2}v}\right)^2+\frac{1}{\left(\frac{\pi}{2}v\right)^2}-\frac{1}{\sin^2\left(\frac{\pi}{2}v\right)}\right]
\label{eq:eq14}
\end{equation}

Solving numerically the equation $(W^H_n)^{''}(v,h)=0$ we can find at which 
$v_0$ the second derivative becomes negative. For $n=3$~we find $v_0=0.6613$ that is a bit closer to $v=0$ than the cubic spline reference value: $2/3$. In Table 3 we give the exact locus of $v_0$~for some fiducial values of $n$. As we can see the higher the exponent $n$ is the closer $v_0$ moves to the origin, making the pairing-instability more unlikely to happen. 

Another interesting feature of $W^H_n$~is that the set is infinitely derivable, with continuous and well-behaved derivatives, while the second derivative of the cubic spline kernel displays several non-derivable points, as depicted in Fig. 4. This may have relevance for those physical magnitudes that rely on second or even higher derivatives. Of course that shortcoming is not shared by higher order splines, such $M_6$~for instance, which have well behaved second and higher order derivatives. In the frequent case of heat diffusion by conduction the standard SPH implementation avoids the calculation of the second derivative by reducing the problem to an integral expression which uses only first derivatives. Nevertheless even with that formulation the smoothness of the second derivative may be of interest because the integral approach involves a balance among the first spatial derivative calculated  through the neighbours of a given particle. Such integral approach is somehow equivalent to perform second derivatives although considerably less noisy.

\subsection{Discrete disordered systems} 

 Interpolation in discrete systems is limited by the presence of 
numerical noise. 
Before attempting to use any particular kernel one has to be sure that the 
local averaged properties of the physical system are not too much distorted by 
its presence so that the noise is not taking over 
the dynamical evolution. Although to really understand the role of the disorder it would be necessary to solve the hydrodynamical evolution of the fluid using 
different levels of resolution, some insight can be gained by studying simple 
cases of interpolation in static lattices. Thus, to complete the study of the 
mathematical features of the 
proposed family of kernels, $W_n^H$, we have extended the 
analysis given in the previous section 
to two-dimensional disordered systems. An uniform grid with N=57600 
particles was built in a box with the masses of the particles conveniently 
crafted  to 
reproduce the density profile of a Gaussian surface. As before the density jump 
across the bell was set to four, but this time the characteristic width of 
the Gaussian was taken large enough to ensure that its profile was 
well resolved by the SPH. The density was calculated using the 
standard SPH equation expressing mass-conservation:

\begin{equation}
\rho_i=\sum_{j=1}^{n_{nb}} m_j W_{ij}(\vert\mathbf r_i-\mathbf r_j\vert,h)
\end{equation}

\noindent
where $m_j$~is the mass of the j-particle and $n_{nb}$~is the number of 
neighbors 
within the compact support.  The number of neighbors was set constant to  
$n_{nb}=43$, 
meaning a ratio $\Delta d/h=0.305$~where $\Delta d$~is the typical 
interparticle separation.   
A level of noise was seeded by randomly re-settling  
these particles so that their final position was within a 
 5\% radius from its initial location at the lattice. 

As can be seen in Fig. 5 the density profile of the Gaussian is not so much 
altered. Because the bell is wide the correct density jump, $\rho_{max}/\rho_0=4$, is 
reached 
with independence of the kernel. As expected the case with n=3 led to a better filtering of the 
noise than n=5 but the differences are not large. Similarly there are not 
serious differences between the 
cubic-spline $M_4$~and quintic $M_6$. Things are different when the gradient of 
density is computed using the SPH expression: 

\begin{equation}
(\mathbf{\nabla_r{\rho}})_i=\sum_{j=1}^{n_{nb}} m_j \mathbf\nabla_r W_{ij}(\vert\mathbf r_i-\mathbf r_j\vert,h)
\end{equation}

\noindent
where $\mathbf\nabla_r$~means the radial component of the gradient. The results are summarized in Fig. 6, where it is evident that disorder has an important 
influence on the first derivative. Even though, on average, the  
first derivative still keeps with the original profile, it becomes blur showing 
an important dispersion, especially close to the peak at $r\simeq 10$~cm. However, low-order interpolators $M_4$~and $W_3^H$~
do show a much lesser dispersion than the higher-order kernels $M_6$~and 
$W_5^H$, 
as expected. Fig. 6 also suggests that the similarity between the pairs  
$M_4, W_3^H$~and $M_6, W_5^H$~kernels shown in Figs. 1 and 4 
still holds 
in disordered systems, being  good enough as to make 
them exchangeable in calculations which use a limited number of particles. 
A quantitative idea of the dispersion is given in Fig. 7 which shows the 
profile of the standard deviation $\sigma$~of the density gradient along 
the box as 
a function of the kernel index $n$. As we can see the dispersion 
becomes more important as $n$~rises. However the trend is not monotonic being 
more accentuated for large values of $n$. 

It has to be stressed that in discrete systems the effect of 
increasing $n$~in $W_n^H$~while keeping the 
number of neighbors, $n_{nb}$, constant would somehow be equivalent to reduce 
$n_{nb}$~leaving the index $n$~unaltered. For example a similar level of noise 
to that shown by $n=5$~can be obtained using $n=3$~but with 
$n_{nb}^{'}\simeq\frac{2}{3}n_{nb}$. Nevertheless it is risky to reduce 
too much 
the number of neighbors because any numerical fluctuation could have a large impact on the smoothed variables.

\section{Numerical simulations}

The discussion given in Section 2 refers basically to the ideal mathematical properties of the $W^H_n$~family of interpolators. Unfortunately, as stated at  
the end of the previous section, much 
of these properties are partially lost in practical applications because of the presence of numerical noise. In current hydrodynamical calculations the physical system is decomposed in a finite number of {\sl particles} with mass $m_i$~and integrals such those given by Eq. (\ref{eq:eq8}) or Eq. (\ref{eq:eq11}) are calculated through a summatory which involves the  neighbours of a given particle. Therefore some level of noise is unavoidable in SPH. In practice a good kernel interpolator should give reliable values for the smoothed variables at low computational cost and keeping the noise at low enough level to not interfere with the simulation. It is well known that slender kernels are better interpolators but they also generates more noise, \cite{fq96} (see also Fig. 6). Thus, despite their good continuum features, choosing a too large $n$-value in Eq. (\ref{eq:eq3}) could bring more problems that solutions unless a large number of neighbours is used (but in that case the main advantage of increasing $n$~is lost and the computational burden rises). In general the selection of the more suitable kernel is dependent to the particular physical situation we want to simulate and even to the available computational resources.  

In this section we describe a couple of standard tests carried out in cartesian 2D aimed at exploring the behaviour of the $W^H_n$~kernels when a limited number of particles is used to describe the fluid. The main goal is, however, to 
provide {\sl practical examples} about the advantages of using harmonic kernels with different index $n$~during the simulation. The first example refers to the propagation of a thermal wave in an homogeneous medium. In this case the setting of the initial conditions has a large impact in evolution of the wave. If stiff initial conditions for the thermal profile are imposed we will show that the use of a kernel with large index $n$~in the energy equation improves the quality of the simulation. The second example deals with the numerical simulation of a point-like explosion in an homogeneous environment, usually referred as the Sedov-test in the literature. In this case the use of high-peaked kernels in the shock front area and, conversely, low-peaked ones in the rarefaction tail leads to a better energy conservation.   

To perform these tests we used a 2D cartesian SPH code with temporal and spatially variable smoothing length. The SPH equations used in the tests calculations are the mass, momentum and energy conservation written in 
their most common formulation, \cite{m92}.  

\begin{equation}
\rho_i=\sum_j m_j W_{ij}(\vert\mathbf r_i-\mathbf r_j\vert,h_i)
\end{equation}

\begin{equation}
\mathbf{a}_i=-\sum_j m_j \left(\frac{P_i}{\rho_i^2}+\frac{P_j}{\rho_j^2}+q_{ij}\right) 
\mathbf{\nabla_i \widetilde W_{ij}}(\vert\mathbf r_i-\mathbf r_j\vert, h_i, h_j)
\label{eq:eq22}
\end{equation}

\begin{eqnarray}
\frac{du_i}{dt}=\frac{P_i}{\rho_i^2}\sum_j m_j \mathbf{v}_{ij}\cdot \mathbf{\nabla \widetilde W_{ij}}(\vert\mathbf r_i-\mathbf r_j\vert, h_i,h_j)\cr\cr
+\frac{1}{2} \sum_j m_j q_{ij} \mathbf{v}_{ij}\cdot \mathbf{\nabla_i \widetilde W_{ij}}(\vert\mathbf r_i-\mathbf r_j\vert, h_i, h_j)
\label{eq:eq23}
\end{eqnarray}

where $\mathbf{v}_{ij}=\mathbf{v}_i-\mathbf{v}_j$, and the other symbols have their usual meaning. The kernel $\widetilde W_{ij}(h_i, h_j)=0.5~ [W_{ij}(h_i)+
W_{ij}(h_j)]$~is symmetric with respect any pair of particles in order to ensure the 
exact conservation of momentum. 
The term labeled as $q_{ij}$~is the artificial viscosity term defined as, 

\begin{equation}
q_{ij}=\cases 
{\frac{-\alpha\bar c_{ij}\mu_{ij}+\beta{\mu^2_{ij}}}
{\bar\rho_{ij}}\qquad\qquad\qquad \mathbf {v_{ij}}\cdot \mathbf {r_{ij}}<0\cr\cr 
0\qquad\qquad\qquad\qquad\qquad\quad \mathbf {v_{ij}}\cdot \mathbf {r_{ij}}>0\cr} 
\label{eq:eq24}
\end{equation}

and, 

\begin{equation}
\mu_{ij}=\frac{\ell~\mathbf{v}_{ij}\cdot~\mathbf{r}_{ij}}{r^2_{ij}+\nu^2}
\label{eq:eq25}
\end{equation}

here $\ell\simeq h$ is a characteristic length which controls the width of the 
shock, $\nu=0.1h$~helps to avoid divergences when $r_{ij}\rightarrow 0$ and the remaining symbols have their usual meaning. The value of the parameters $\alpha, \beta$~were set to $1$~and $2$~respectively in all simulations. 

These equations were 
completed with the diffusive heat transfer equation when necessary:

\begin{equation}
\left(\frac{du}{dt}\right)_i=-\sum_j m_j \frac{(q_i+q_j)(u_i-u_j)(\mathbf{{r}}_{ij}\cdot{\mathbf{\nabla}\widetilde  W_{ij}})}{\bar\rho_{ij}(r^2_{ij}+\nu^2)}
\label{eq:eq17}
\end{equation}

where $q=\kappa/(\rho c_v)$, (being $\kappa$~the conductivity coefficient and 
$c_v$~the specific heat), $\bar\rho_{ij}=0.5(\rho_i+\rho_j)$, $\mathbf{r}_{ij}=\mathbf{r}_i-\mathbf{r}_j$~and $\nu=0.1h$~is a term which avoids divergences when $r_{ij}\rightarrow 0$. The expression above has the peculiarity that only first derivatives of $W^H_n$~need to be computed at each step. 

Contributions from temporal and spatial derivatives of the smoothing-length 
parameter were neglected for simplicity. While that approach does not pose a 
problem for the 
first test, because the thermal wave is supposed to propagate through an
 homogeneous medium of static particles, it is not evident its validity for 
the Sedov explosion calculation. It has been shown that the inclusion of the smoothing-length 
derivatives often improves the energy conservation \cite{n94}, although 
such enhancement seems only relevant for a restricted class of problems (for 
example the head on collision of two polytropes). Keeping in mind that the 
main goal of the Sedov test shown below is to make a comparative analysis 
among different kernels rather than to solve the problem with great accuracy, 
we have preferred not to include the derivatives of $h(r,t)$~in the equations.

Motion equations were integrated using a second order centered scheme. A squared lattice with one particle in each node and 1 cm of distance between adjacent neighbours was implemented. The box has 240 cm of side length, hence there are 57600 particles, and uses periodic boundary conditions. With only one 
exception (the mixed case in the Sedov test), the value of $h(r,t)$~
self-adapts to keep a constant number of particles, $n_{nb}=43$, within the 
kernel radius. That number of neighbors means a ratio $\Delta d/h=
0.305$~being $\Delta d$~the interparticle separation. All particles have the same mass, adjusted to obtain an uniform density profile with $\rho_0=1$~g.cm$^{-3}$, and obey a perfect gas EOS, $P=(\gamma-1)\rho u$~with $\gamma=5/3$, being $u$~the specific internal energy.   

\subsection{Thermal wave}

In this test we follow the propagation of a thermal wave moving through an homogeneous medium with constant density. During the calculation we obliged the particles to be at rest so that the evolution of the wave was determined by evolving only the energy equation. For this particular problem the main difficulty  
relies in the accurate resolution of the diffusive heat transfer 
equation, Eq. (25) below. Although the numerical resolution of that equation 
usually demands the calculation of second derivatives there is a clever 
formulation in SPH, \cite{b85} that reduces the problem to first derivatives. 
Nevertheless such procedure involves the balance between the first derivative 
of the kernel evaluated in discrete points inside its compact support 
area, which is equivalent to calculate its second derivative.   
As the most common used interpolator $M_4$~does not have a well behaved second 
derivative, Fig. 4, it can be instructive to compare the evolution of 
the thermal wave calculated  using $M_4$~to 
 that using the $W_n^H$~family and to $M_6$. Another point of interest of 
this test relies in the hard initial conditions, which attempts to represent a
 thermal discontinuity. As in the case of the shock waves the multidimensional 
hydrocodes have also difficulties to handle thermal discontinuities which 
have to be artificially enlarged to the resolution of the code. An extreme 
example 
of a very sharp steady thermal wave (with a thickness between  
$10^{-2}-1$~cm) is the 
precursor thermal wave which induces 
the propagation of a self-sustained nuclear flame in Type Ia Supernovae 
explosions \cite{h00}. For these cases it could be very useful to use 
a highly condensed kernel, $n\simeq 5-6$, in $W_n^H$ to handle the heat 
diffusion and the standard, $n=3$, for the remaining equations.

For a given initial conditions we have carried 
out five runs using the kernels $M_4$, $M_6$, $W^H_3$,$W_5^H$~and $W_6^H$.
 
\subsubsection{Initial model}
We have followed the recipe by \cite{jsd04} (see there for details on numerical implementation of conduction in SPH). Initializing the internal energy, following the two-dimensional Green's function, the system is exactly in a state that is a solution of the conduction problem for an initial $\delta$-function:

\begin{equation}
u(x,y,t)=\frac{A}{4\pi \alpha t}\exp{\left(-\frac{x^2+y^2}{4\alpha t}\right)}+u_0
\label{eq:eq15}
\end{equation}

where $A=10^5$ erg.cm$^2$/g, $\alpha =1$ cm$^2$/s is the thermal diffusivity 
and  $u_0=10^3$ erg/g. For a given elapsed time $t$~Eq. (\ref{eq:eq15}) provides the precise profile of the thermal wave emerging from the initial discontinuity. At $t=0.25 s$~the width of the signal becomes equal to the smoothing length $h$. We take the thermal profile at that time as the initial conditions of our simulations.  That setting is in fact quite unfavorable for the numerical approach but it is adequate to show the advantages of using a kernel with a high $n$-value in $W^H_n$.

\subsubsection{Results}

The evolution of a thermal wave moving inside a static media is basically determined by the mass conservation equation, Eq. (18), and the energy equation:  

\begin{equation}
\frac{du}{dt}=\frac{1}{\rho}{\mathbf{\nabla}} \cdot \left(\kappa{\mathbf{\nabla} T}\right)
\label{eq:eq16}
\end{equation}

where $u$~is the specific internal energy and $\kappa$~is the coefficient of conductivity. The SPH version of Eq. (25) was  Eq. (23) given above. 

The rate of change of the thermal energy content can be obtained by deriving Eq. (24) with respect time,

 \begin{equation}
\frac{du}{dt}\left(r,t\right)=\frac{A}{4\pi \alpha t^2}\exp{\left(-\frac{r^2}{4v\alpha t}\right)}\left[\frac{r^2}{4 \alpha t}-1\right]
\label{eq:eq18}
\end{equation}

The evolution of the specific internal energy profile at t=0.30 s (thus, close to our starting time t=$0.25$s), t=1 s and t=5 s is shown in Fig. 8. As the initial state at t=0.25 s was the same for the five kernels the $u$-profile at the upper-left picture does not show significant differences. At t=1 s (middle picture) the thermal signal has moved to the right and the peak at the origin has decreased. It can be seen that neither the $n=6$~nor the $n=3$~case correctly match the analytical prediction. However $W^H_6$~makes a better job than $W^H_3$. When the elapsed time was  t=5 s the profile has already become smooth enough that the evolution was independent of the kernel index $n$. Nevertheless the differences close to the initial discontinuity (see the profiles for $r\le 2.5$~cm) still remains.  

The differences in the outcome of the simulations can be more easily analyzed if we monitor the temporal evolution of the maximum value of the derivative of the internal energy and its position, 
  
\begin {equation}
\frac{du}{dt}\bigg\vert_{max}\left(t\right)=\frac{A}{4\pi \alpha \e^2}t^{-2}
\label{eq:eq19}
\end{equation}

\begin{equation}
r_{max}=2\left(2\alpha t\right)^\frac{1}{2}
\label{eq:eq20}
\end{equation}

The evolution of $(du/dt)_{max}$~is depicted in  Fig. 9. Again none of the kernels were able to reproduce the analytical value. Such difficulty is of course caused by the sharp initial profile imposed to the internal energy. Nevertheless it can be seen that $W_6^H$ is closer to the analytical solution than $W_3^H$~for the firsts stages of the evolution, when the width of the peak of $du/dt$ is of the order of the smoothing length.
The use of $W_6^H$ is about a 30\% more accurate than $W_3^H$ when the system is in a such disadvantageous state for numerical simulations. 
As the system evolves the peak in $du/dt$ widens becoming much larger than the smoothing length so SPH can resolve it accurately and both values of index $n$ give the same results. 
The case with $n=5$~provides a better approximation than that of $n=3$~although 
 slightly worse than $n=6$. An inspection of Figs. 8 and 9 also indicate that 
there are 
not significative differences 
between the evolution of the thermal wave calculated using the polynomial 
functions $M_4$, $M_6$~and harmonic interpolators $W_3^H$, $W_5^H$.

\subsection{Sedov explosion}

In the Sedov test the evolution of a shock wave front is studied as it propagates in a homogeneous medium. The problem of an intense explosion in a gas is a standard test for hydrocodes and has some relevance in astrophysics, where is common to find strong shocks in many scenarios involving fluid motions at high velocity. The theoretical solution was found by L.I. Sedov applying self-similar methods and dimensional analysis for different geometries and values of $\gamma$, \cite{s59}. 
In its simplest formulation the Sedov problem has an initially cold gas at rest. At 
t=0 s there is a point explosion at the origin which in \cite{s59} was treated 
as an instantaneous release of energy at the origin and assumed that the 
background material through which the expanding gas sweeps behaves as a perfect fluid with $P=(\gamma-1)\rho u$. The most remarkable feature of this problem 
is that it leads to exact, although algebraically complicated, analytical 
expressions for the fluid variables. Unfortunately it is not easy to exactly 
simulate the evolution of the blast wave in more than one dimension. As in the 
thermal wave test the shock front was too sharp to be resolved by the hydrocode. 
In 
the case of SPH the artificial viscosity smears the shock over 2-3 times the 
smoothing-length. As a consequence the density jump across the shock front 
is always lesser that the factor four predicted by the theory for $\gamma=5/3$.
Thus, resolution issues are here crucial not only to resolve the peak but also 
to reproduce the correct postshock variables downstream and the structure of 
the rarefied tail close to the origin. The use of adaptive kernels can 
greatly help to handle with this problem. In this respect the $W_n^H$~family 
of interpolators adds an extra degree of freedom to the scheme which, combined
with a clever use of the adaptive smoothing-length $h(r,t)$~is able to bring 
a better approach to the Sedov problem.

We have carried out six calculations with the same initial conditions in order to analyze the influence of using kernels with different index $n$~in the outcome of the explosion. First we have simulated the evolution of the blast wave for $W_3^H$, $W_5^H$~$W_6^H$~and $M_4$, $M_6$ comparing the results with the 
analytical profile. In our second simulation we have taken an adaptive index $n$ in $W_n^H$~which changes according to the compressional state of the material. In that case a clear improvement of the energy conservation was seen with respect to the calculation that relied in the cubic spline. An alternative to Eq. (20) which 
ensures the mechanical energy conservation is to consider the evolution of the 
thermokinetic, $\frac{1}{2} v^2+u$, energy \cite{m05}. Nevertheless solving the 
thermokinetic equation does not mean that the internal energy is better evolved
 than using Eq. (20), especially if there are sources or sinks of energy.

\subsubsection{Initial model}
In order to generate a shock wave, an amount of particles (about 5\% of the total number of particles) that lie inside a box-centered Gaussian, had their pressure artificially raised. To smooth the initial discontinuity we take an initial pressure step that decays as a Gaussian function,  

\begin{equation}
P(r)=P_2+\left(P_1-P_2\right)\exp{\frac{-r^2}{{\sigma}^2}}
\label{eq:eq21}
\end{equation}

where $P_1$ and $P_2$ are the pressures in both zones -left and right around the pressure step- and $\sigma$ sets the width of the pressure decay. In our simulations $P_1=10^4$ dyn/cm$^2$ and $P_2=1$ dyn/cm$^2$, to assure an internal energy reservoir big enough to feed the formation of the shock wave, and $\sigma^2=16$ cm$^2$, which smoothes the pressure step over about two times the smoothing length. Again the initial profile is so sharp that SPH finds some difficulty to track the blast wave. In particular the conservation of energy is not longer satisfactory during the transient period until the self similar wave appears. However the comparison between models with different value of $n$~is meaningful. We have taken the most common form of the mass conservation, momentum and energy equations,  
\subsubsection{Results}

Six different calculations characterized by identical initial conditions but allowing variations in the smoothing kernel were carried out. Shortly after the induction of the initial explosion a steady self-similar state ensued. In all the cases we found that the evolution of the blast wave matched well the analytical results after the self-similar state was achieved. There were, however, interesting differences among the models which were caused by the type of kernel used in each calculation. The density profiles as a function of the normalized distance during the self-similar state are depicted in Fig. 10. As expected $M_4$~and $W_3^H$~led to an almost identical profiles but, interestingly, the harmonic kernel gave a slightly better energy conservation, as it can be seen in Fig. 11. The simulation which used the  $W_6^H$~kernel led to a large enhancement in the energy conservation, owing to  the greater weight imprinted by the nearest neighbors, which makes more compatible  momentum and  energy equations, Eqs. (\ref{eq:eq22}) and  (\ref{eq:eq23}). Nevertheless the density profile during the calculation with $n=6$~was not so well defined as in the previous cases, as it can be seen in the fifth snapshot of Fig. 10. There is more noise, inherited from the initial distribution of particles in a square lattice. Still the dispersion is affecting a reduced number of particles and the profile also follows the 
analytical curve. The source of the noise can be seen in Fig. 12 which 
depicts several snapshots of the density evolution for the cases $n=3$~and 
$n=6$. In the first row, obtained using $n=3$, the symmetry was conserved to a 
high degree whereas it was not so well preserved for $n=6$~in the second row, 
which shows 
irregularities at $0, \pi/2, \pi$~and $3\pi/2$~rad, even at the first 
snapshot. Smaller irregularities are also apparent in the last picture at 
$\pi/4,~ 3\pi/4,~ 5\pi/4$~and $7\pi/4$~rad. These irregularities come from 
the small instabilities seeded by the rectangular lattice used to set the 
simulations. That  geometry imposes preferred directions for strain 
propagation, a phenomenon known as hourglass instability. The more efficient 
smoothing given by low-order kernels damped the grow of that instability from   
the beginning in the same way that they made a more efficient  
filtering of the random 
noise depicted in  Fig. 6.
The evolution for $n=5$~is close to that of $n=6$~although a little less noisy. As in previous calculations the pair   
$W_5^H$~harmonic kernel and $M_6$~spline led to a very similar, if not identical, evolution.  
Only  the conservation of energy seems to be slightly better 
for $W_5^H$. 

These results suggest that the properties of low and high $n$-indexed kernels could be combined to improve the quality of the simulation without introducing spurious numerical noise. A way to do that is to allow the smoothing length to take values so that the number of neighbours, $n_{nb}$, of each particle remains in a prescribed range rather than oblige them to be constant, as in the calculations above. Close to the shock edge the number of neighbours tends to rise whereas the opposite is true along the rarefaction wave. More (less) neighbours means 
less (more) numerical noise which can be compensated taking a 
high (low) $n$~value in $W_n^H$. Thus, in our last calculation we 
allowed $n_{nb}$~to stay in the range $20\leq n_{nb}\leq 80$~and found the 
exponent $n$~from the expression: $n=2.88539\ln (n_{nb})-6.6438$~
which gives n=2 for $n_{nb}=20$ and n=6 for $n_{nb}=80$. Such range in $n_{nb}$~led to an average in the number of neighbours $\bar{n}_{nb}\simeq 44$~during the run, similar to that taken to compute the unmixed cases above. Knowing the particular value of $n$, Eq. (\ref{eq:eq6}) allows to compute the normalization constant. For a meaningful comparison with the previous simulations  the parameter $\ell$~used in the artificial viscosity 
coefficient given by Eq. (\ref{eq:eq25}) was kept equal to that taken to 
compute 
the $n=3, 5, 6$~and $M_4, M_6$~cases. The resulting density profile is shown in the sixth snapshot of Fig. 10 and the evolution in the energy conservation is depicted in Fig. 11. As we can see conservation of energy is half-way between that of cases $n=3$~and $n=6$~with the advantage that the numerical noise which plagued the $n=6$~case has vanished. In addition the high-density part of the profile is even a little better reproduced than for the $M_4$~ or for $n=3$~case. However the fit of the low-density region behind the wave, although satisfactory, is not as good as for the $n=3$~case.

\section{Discussion and Conclusions}

A one-parameter family of interpolating kernels with compact support based on the harmonic-like functions, $W^H_n$, defined by Eq. (\ref{eq:eq3}) have been analyzed and checked in the context of the smoothed particle hydrodynamics method. Formally the widely used polynomial interpolators are linked to the proposed kernels via the Fourier transform of a function similar to $W^H_n$. We have found that the set of functions defined by Eq. (\ref{eq:eq3}) also displays good enough mathematical features as to deserve being considered smoothing kernels by themselves. Using $W^H_n$~has several interesting advantages: a) if the exponent $n$~ 
 is conveniently chosen these interpolators are able to mimic with great 
accuracy various of the most common kernels used so far in SPH studies. In 
particular the cubic and quintic spline kernels $M_4$~and $M_6$~are well 
reproduced by taking $n=3$~and $n=5$~in Eq. (3). Even the truncated Gaussian 
kernel is 
reasonably fit for $n=2$. Such equivalences seem to be robust in 
the light of the two realistic test cases analyzed in Section 3,
 b) in the limit of a very large number of particles the use of functions with $n>3$~improves the resolution when strong gradients are present (Section 2), c) Unless the cubic spline, whose second derivative is not smooth in several points, the $W^H_n$~has well-behaved second derivatives allowing the set to be derived many times, d) it adds even more flexibility to the SPH technique because the quality of the interpolation can be selected by simply varying the exponent $n$~in equation 3. Changing the exponent $n$~is really straightforward and does not introduce any computational overload in the numerical scheme. 

In current calculations, where a moderate number of particles are put into the system, the choice of kernels with a large exponent $n$~is limited by the numerical noise. Therefore the use of high-peaked interpolators should be restricted 
to special circumstances. It has been shown in Section 2 that the 
improvement in resolution is not monotonic as $n$~rises: above $n=6$~it is  
hard to achieve a much better resolution. Just the opposite is true for the 
noise because it grows faster for $n>6$. Thus, it is advisable 
to restrict the range of the exponent $n$~to the interval $2\le n\le 6$~in practical 
applications.

Another question related to the number of particles is that either the cubic or 
the quintic spline kernels can also have their resolution increased by decreasing the smoothing-length parameter $h$. Reducing $h$~has a similar effect as increasing $n$~in Eq. (\ref{eq:eq3}) because there is an enhancement of the resolution accompanied by an increase of the numerical noise due to the reduction in the number of neighbours of the particle. That enhancement in the resolution can 
be used to devise an adaptive kernel scheme which has proven useful to handle
 shocks \cite{s06}. 
Nevertheless varying the resolution 
by altering $h$~in the cubic spline is not as good as to change $n$~in $W^H_n$~for two reasons. First, a high-peaked kernel is still able to count particles which are farther than those seen by a low-peaked one with comparable resolution but lesser $h$. In some circumstances the statistical weight of these particles could be high enough to influence the average. The second reason is simply that to control both the resolution and the noise it is also better to have two parameters to tune: the size of $h$, which sets the number of neighbours of a given particle, and the exponent $n$~in Eq. (\ref{eq:eq3}).   

The only drawback of the $W^H_n$~family is that a trigonometric function has to be calculated every time the kernel is invoked during the simulation. That leads to a computational overload with respect to the evaluation of the cubic spline. More specifically, if we do a large number of calls to the cubic spline kernel and to $W^H_3$~with random generated arguments, the ratio in CPU time is 
roughly a factor 2.5 favourable to the polynomial kernel. Nonetheless 
in a real 
hydrodynamical calculation such factor is diluted by the rest of the 
computations especially if gravity is present or the numerical algorithm 
includes complex physics. 

As a general rule we propose to implement the kernels $W^H_n$~in the SPH equations leaving the index $n$~free. In normal conditions, i.e. when there are not strong gradients in the physical magnitudes, we should take $n=3$~because for that value the kernel behaves as the well checked cubic spline. In several circumstances, though, it could be wise to turn the value of the exponent $n$~to a different value. It could be taken, for instance,  $n=3$~in all the SPH equations except in that one concerning the conduction transport term given by Eq. (\ref{eq:eq17}) if a thermal discontinuity, a hot wall for example, is present (as suggested in Section 3.1). When using a second order Runge-Kutta type integrators we may want to use a different value of $n$~during the first and the second (centered) integration steps. Another possibility is to allow each particle to carry its own $n$~value as in the Sedov test described in Section 3.2 where the combination of an adaptive $n(\mathbf{r},t)$~and $h(\mathbf{r}, t)$~improved the fit of the density around the peak and led to a better energy conservation. In some cases even the choice of kernels with $n<3$ could be worthwhile to reduce 
the numerical noise.  

Taking into account their easy implementation, smoothness and flexibility, this family of kernels are an alternative to the widely used spline kernels, offering several 
advantages that may help to improve hydrodynamical simulations which use the SPH technique.

\noindent
{\bf Acknowledgements}

The authors want to thank the referees for the constructive comments and 
many suggestions which have contributed to improve the scientific content 
and general presentation of this manuscript. 
This work has been supported by the Spanish MCYT grant AYA2005-08013-C03-01.



\clearpage
\begin{figure}
\includegraphics[scale=1.]{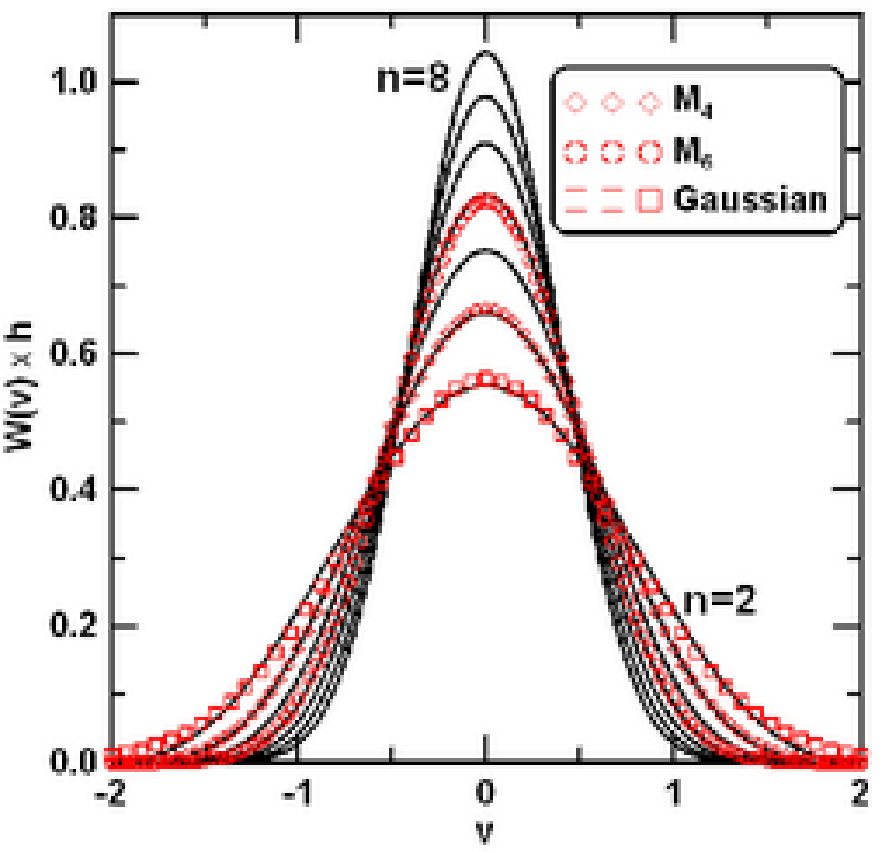}
\caption{Profiles of $W^H_n\times h$~in 1D for several values of n in the range n=2 to n=8. Superposed are also depicted the profiles of various of the most 
common kernels used in SPH: $M_4$~ (cubic-spline), $M_6$~(quintic spline) and 
truncated Gaussian.}
\end{figure}

\clearpage
\begin{figure}
\includegraphics[scale=1.]{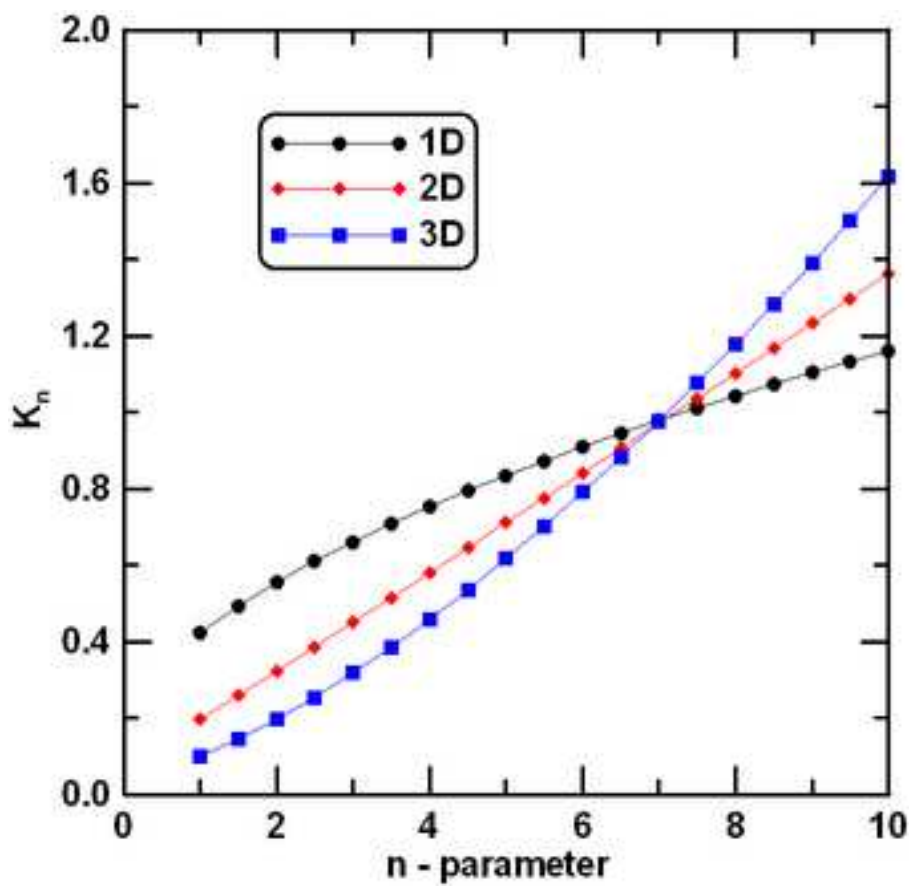}
\caption{Normalization constants $K_n$~of $W^H_n$~for different values of parameter $n$ in one, two and three dimensions.}
\end{figure}

\clearpage
\begin{figure}
\includegraphics[scale=.6]{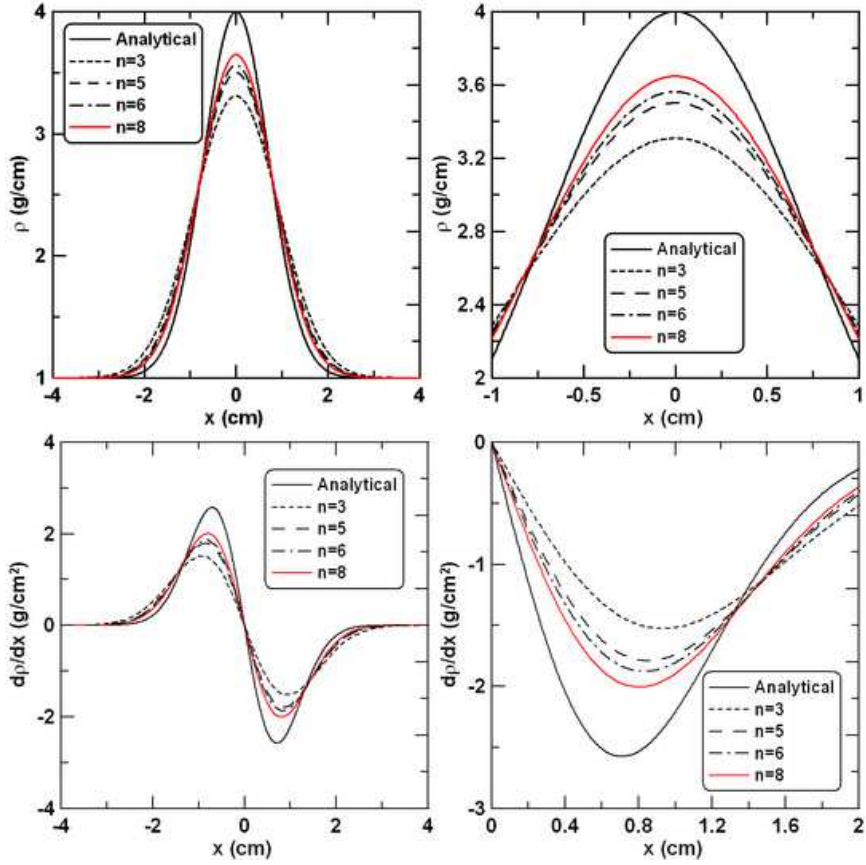}
\caption{Upper-left and right (for a zoom): Ability of the $W^H_n$~set to fit an one-dimensional density distribution with a sharp Gaussian profile. When the characteristic width of the Gaussian is equal to the smoothing-length parameter $h$~a high value of $n$~leads to a better approximation to the analytical density profile. 
Bottom-left and right (for a zoom): The derivative of density is also better 
approached by taking a large $n$. Note that the enhancement in resolution is 
not monotonic as $n$~ rises.}
\end{figure}

\clearpage
\begin{figure}
\includegraphics[scale=.5]{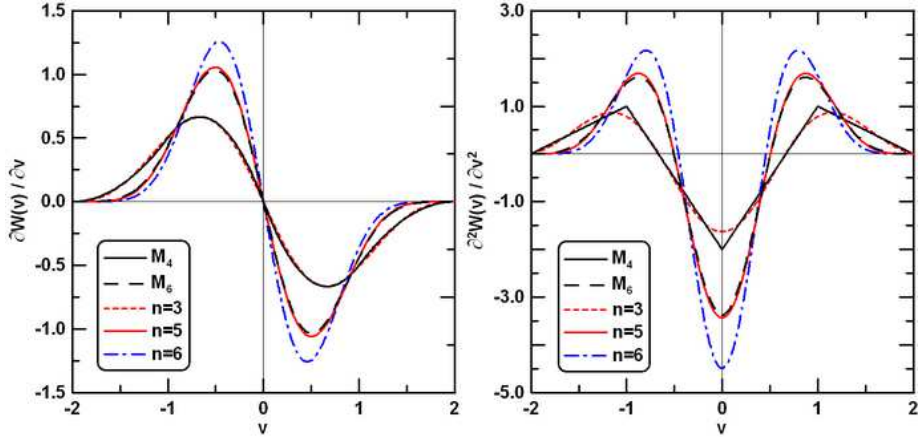}
\caption{First (left) and second (right) derivatives of $W^H_n$~for $n=3,5, 6$~and splines $M_4, M_6$. The first derivative of the pair $M_4, W_3^H$~behaves similar but 
the second derivative profile of $W_3^H$~is smoother. The pair $M_6, W_5^H$~  
shows a good match in both the first and the second derivatives.}
\end{figure}

\clearpage
\begin{figure}
\includegraphics[scale=.6]{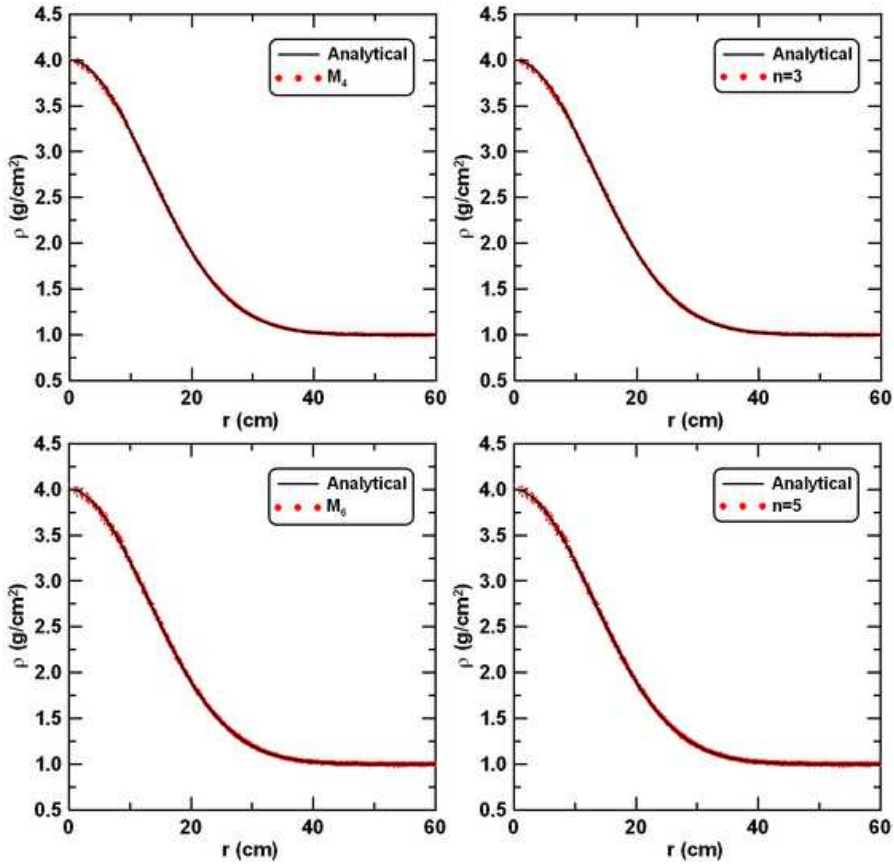}
\caption{Density profile resulting from a two-dimensional distribution of 
slightly 
disordered particles aimed at reproducing a blurred Gaussian surface.  
 Low-order interpolators such $M_4$~or $W_3^H$~(upper-left and right) give 
a more defined profile than the high-order ones $M_6$~and $W_5^H$, but the 
differences 
are not large.}  
\end{figure}

\clearpage
\begin{figure}
\includegraphics[scale=.6]{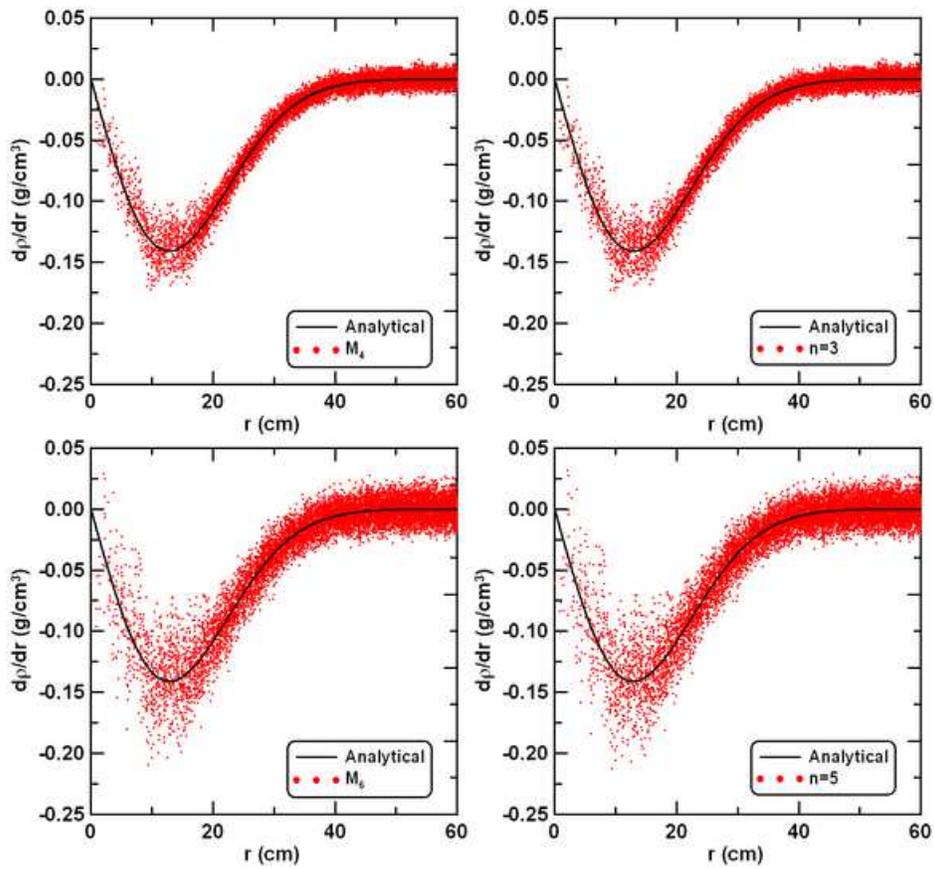}
\caption{Same as Fig. 5 but for the density gradient. Now the imprint of the 
disorder is much larger. Low-order kernels give a more efficient filtering 
of the noise}
\end{figure}

\clearpage
\begin{figure}
\includegraphics[scale=1.]{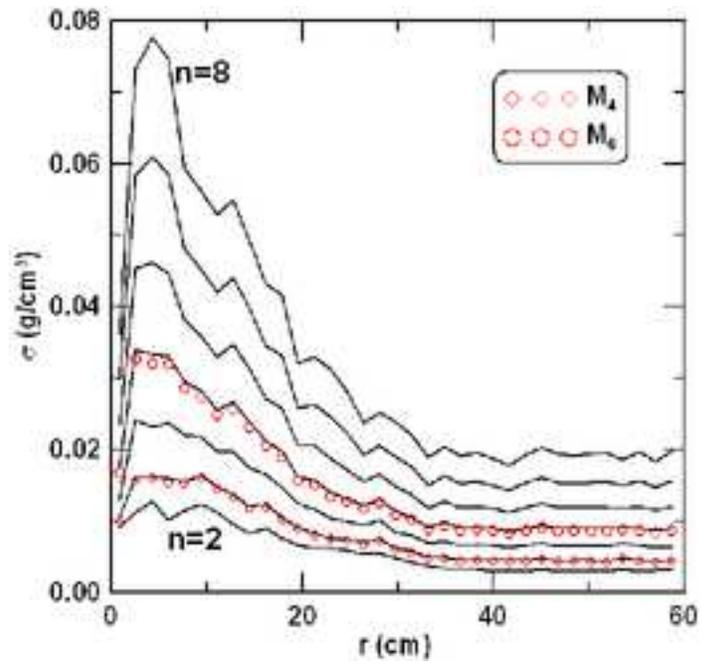}
\caption{Standard deviation of the density gradient profile shown in Fig. 6 
for $n=2$~to $n=8$~and $M_4$, $M_6$. The increase in the standard deviation 
is not monotonic as $n$~rises.}
\end{figure}

\clearpage
\begin{figure}
\includegraphics[scale=.7]{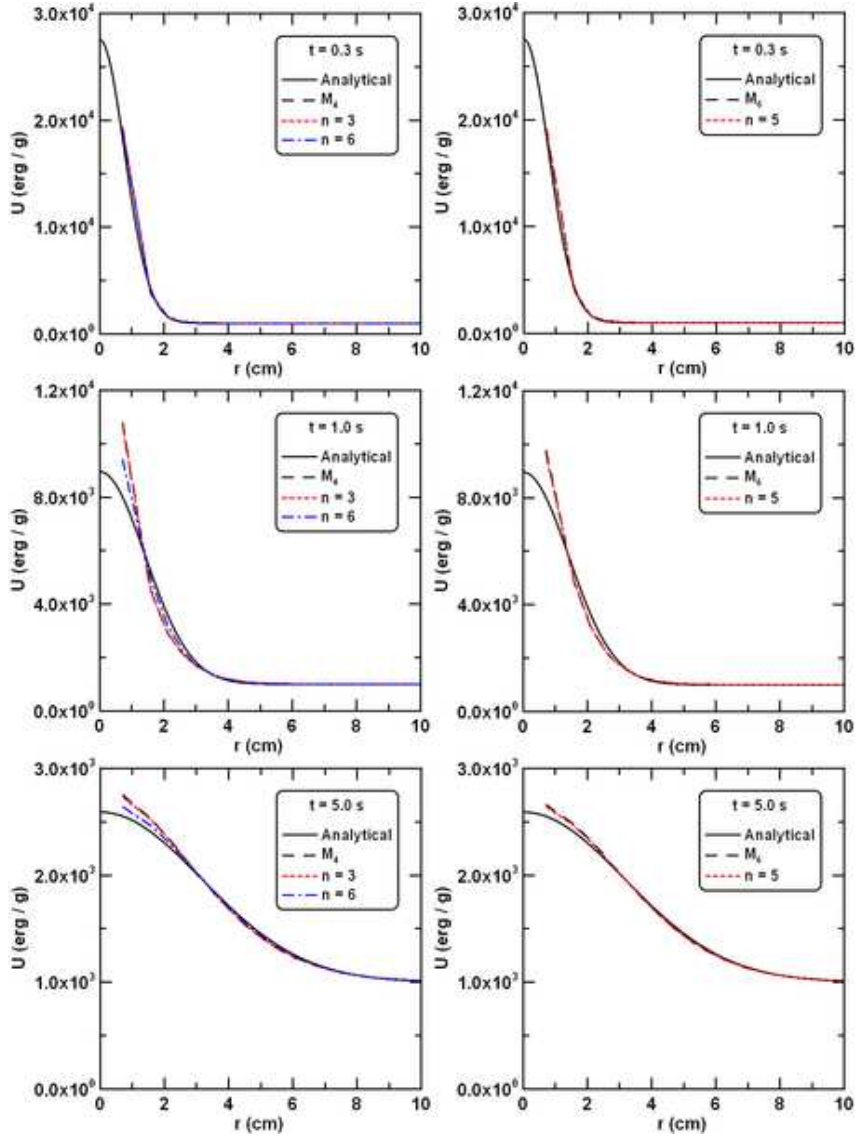}
\caption{Profiles of a thermal wave arising from a discontinuity for several elapsed times. The cubic spline and the $n=3$~case give very similar results.  
The evolution calculated using the quintic spline or $W_5^H$~is also very 
similar. 
The choice of a kernel with a high exponent $n$~in the energy equation improves the  simulation.}
\end{figure}

\clearpage
\begin{figure}
\includegraphics[scale=1.]{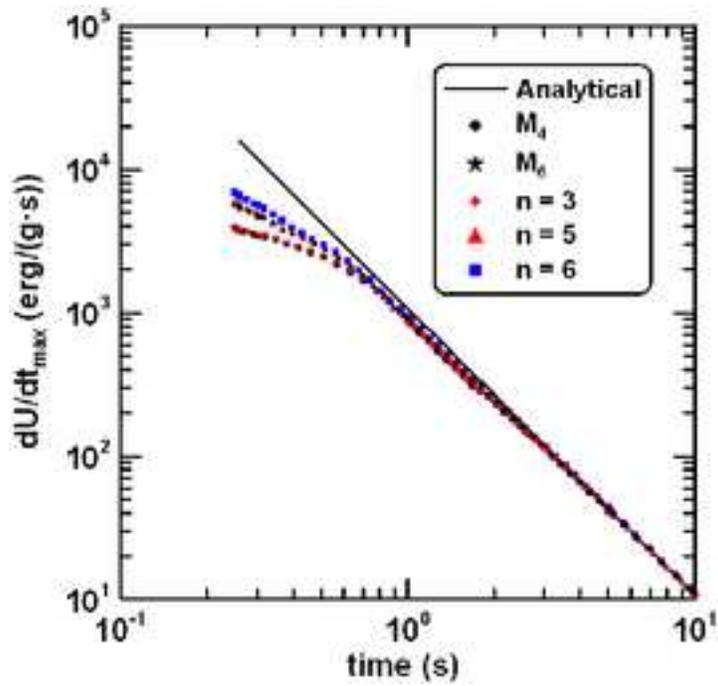}
\caption{Temporal evolution of the maximum flux of heat during the propagation of the 
thermal wave depicted in Fig. 8. After t=1 s the evolution was independent of the particular index $n$~ of the kernel because the thermal profile has become 
quite smooth}
\end{figure}

\clearpage

\clearpage
\begin{figure}
\includegraphics[scale=.7]{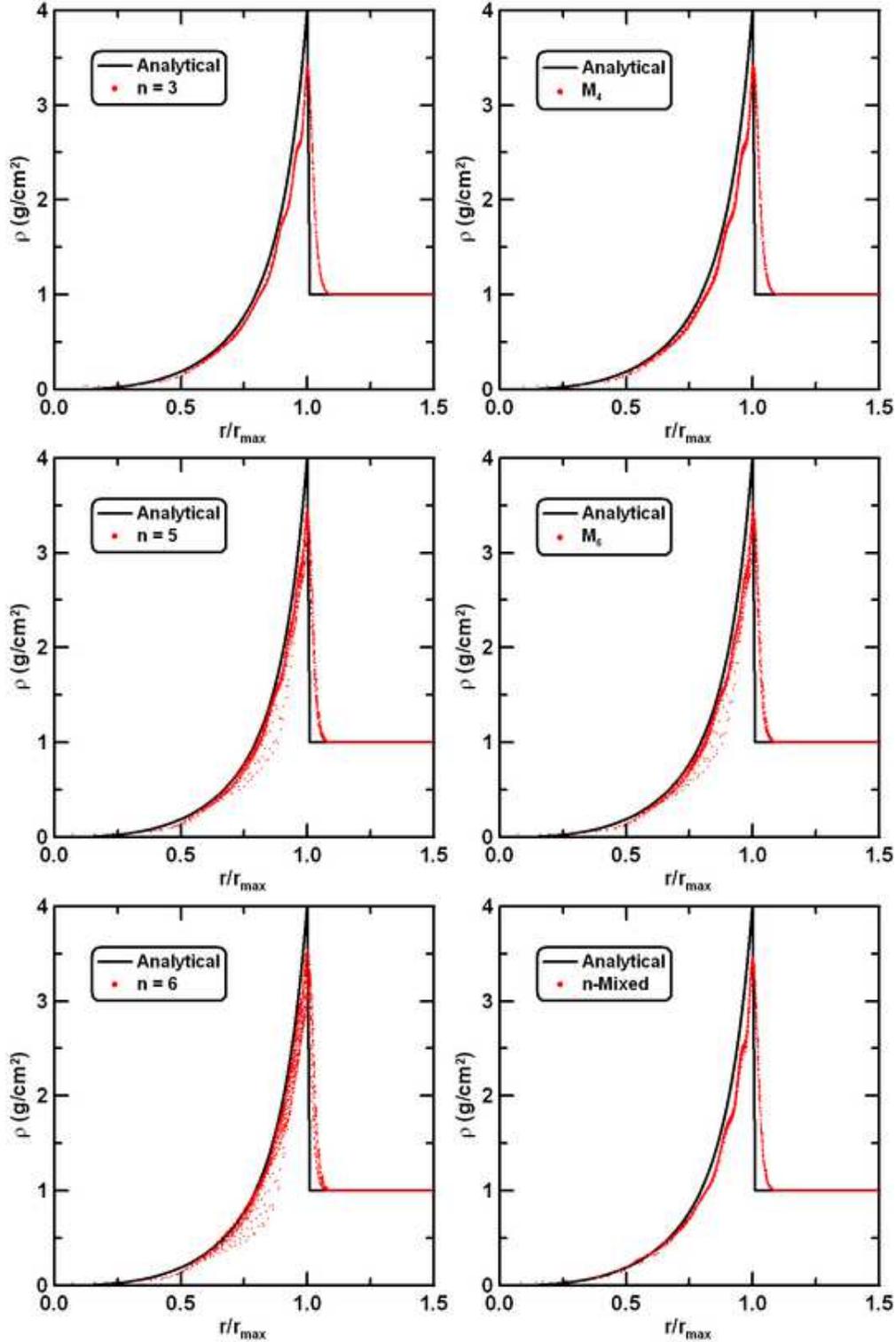}
\caption{Sedov test. Density profiles during the self-similar stage for 
 different choices of the index $n$ in $W^H_n$: $n=3$ (upper-left), $n=5$~(middle-left) and $n=6$ (botton-left). The profiles using $M_4$~(upper-right) and 
$M_6$~(middle-right) are also depicted. The  $n$-mixed case is shown at 
the bottom right. The maximum value of the peaks are $\rho_{max}=3.43$~g.cm$^{-3}$~(cubic-spline), $\rho_{max}=3.42$~g.cm$^{-3}$~($n=3$), $\rho_{max}=3.53$g.cm$^{-3}$~(quintic spline), $\rho_{max}=3.52$~g.cm$^{-3}$~($n=5$),
 $\rho_{max}=3.55$~g.cm$^{-3}$~($n=6$) and $\rho_{max}=3.47$~g.cm$^{-3}$~($n$-mixed).}
\end{figure}

\clearpage
\begin{figure}
\includegraphics[scale=1.]{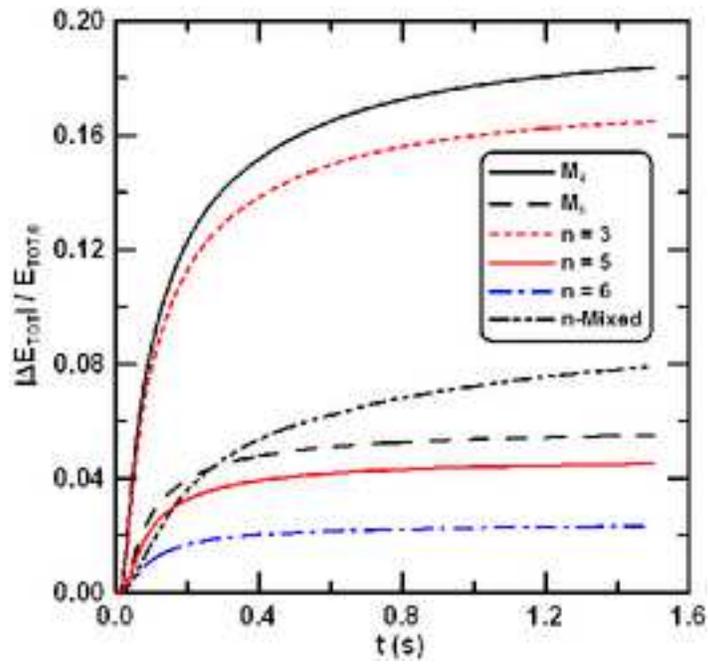}
\caption{Energy conservation during the Sedov test using different 
interpolators. The 
bad conservation especially shown by low-order kernels was due to the hard initial 
conditions.}
\end{figure}

\clearpage
\begin{figure}
\includegraphics[scale=.7]{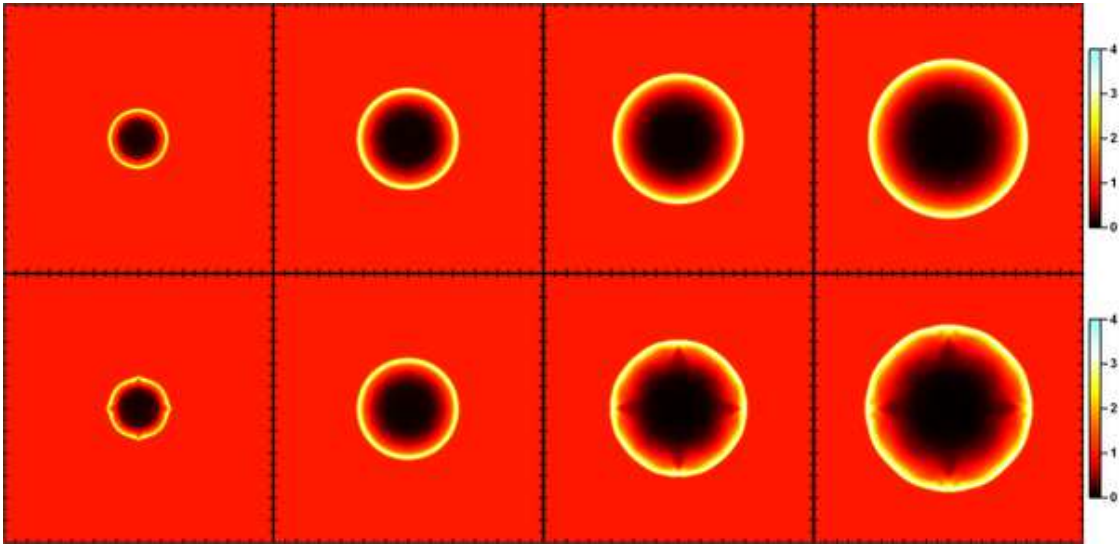}
\caption{Colour-map showing the evolution of density  of a point-like explosion in an homogeneous media (Sedov test) at times (from left to right) t=0.2 s, 0.6 s, 1.0 s and 1.5 s respectively. The upper-row was obtained using $W_3^H$. The lower row is for 
$W_6^H$.
} 
\end{figure}

\clearpage
\begin{table}
\caption{Normalization coefficients $K_n$ for different values of $n$ and dimensions (columns 2,3 and 4). Quotient of integrals $I_2$~and $I_1$~in Eq. (9) (columns 5,6 and 7) related to the maximum value of a sharp Gaussian curve. Factor $(n(I_3/I_1)$~in Eq. (13) (only for 1D, column 8) related to the maximum of the derivative of the Gaussian bell.} 
\vspace{5mm}
\begin{tabular}{|c||c|c|c||c|c|c||c|}
\hline
\multicolumn{4}{|c||}{$K_n$} & \multicolumn{3}{|c||}{$I_2/I_1$} & $n(I_3/I_1)$\\
\hline
n & 1D & 2D & 3D & 1D & 2D & 3D & 1D \\
\hline
\hline
1 & 0.424095 & 0.196350 & 0.098175 & 0.622276 & 0.417078 & 0.297324 & 0.263691 \\
2 & 0.553818 & 0.322194 & 0.196350 & 0.714339 & 0.529116 & 0.404909 & 0.393104 \\
\hline
3 & 0.660203 & 0.450733 & 0.317878 & 0.769870 & 0.605666 & 0.486165 & 0.476948 \\
\hline
4 & 0.752215 & 0.580312 & 0.458918 & 0.807105 & 0.660911 & 0.548647 & 0.53.5367 \\
5 & 0.834354 & 0.710379 & 0.617013 & 0.833859 & 0.702584 & 0.597875 & 0.578325 \\
\hline
6 & 0.909205 & 0.840710 & 0.790450 & 0.854038 & 0.735124 & 0.637552 & 0.611219 \\
\hline
7 & 0.978402 & 0.971197 & 0.977949 & 0.869814 & 0.761234 & 0.670167 & 0.637203 \\
8 & 1.043052 & 1.101785 & 1.178511 & 0.882493 & 0.782649 & 0.697430 & 0.658243 \\
9 & 1.103944 & 1.232440 & 1.391322 & 0.892909 & 0.800531 & 0.720550 & 0.675626 \\
10 & 1.161662 & 1.363143 & 1.615708 & 0.901621 & 0.815690 & 0.740399 & 0.690227 \\
\hline
\end{tabular}
\end{table}

\clearpage
\begin{table}
\caption{Fitting coefficients to the normalization constant $K_n$.}
\vspace{5mm}
\begin{tabular}{|c|c|c|c|}
\hline
 & 1D & 2D & 3D \\
\hline
\hline
$a_0$ & 2.645649$\times 10^{-1}$ & 7.332473$\times 10^{-2}$ & 2.719002$\times 10^{-2}$\\
$a_1$ & 1.824975$\times 10^{-1}$ & 1.196425$\times 10^{-1}$ & 5.469083$\times 10^{-2}$\\
$a_2$ & -2.426267$\times 10^{-2}$ & 3.319287$\times 10^{-3}$ & 1.711166$\times 10^{-2}$\\
$a_3$ & 3.112410$\times 10^{-3}$ & -5.511885$\times 10^{-4}$ & -1.237265$\times 10^{-3}$\\
$a_4$ & -2.404560$\times 10^{-4}$ & 4.828286$\times 10^{-5}$ & 8.193975$\times 10^{-5}$\\
$a_5$ & 8.032609$\times 10^{-6}$ & -1.733766$\times 10^{-6}$ & -2.552696$\times 10^{-6}$\\
\hline
\end{tabular}
\end{table}

\clearpage
\begin{table}
\caption{Critical value $v_0$~at which \newline the second derivative of \newline $W^H_n$~and $M_4, M_6$~becomes negative \newline leading to pairing-instability.}
\vspace{5mm}
\begin{tabular}{|c|c|}
\hline
  Interpolator& $v_0$ \\
\hline
\hline
$M_4$ & 2/3 \\
$M_6$ & 0.5062\\
$W_3^H$ & 0.6613 \\
$W_5^H$&0.5039\\
$W_6^H$ & 0.4582 \\
$W_9^H$ & 0.3718 \\
\hline
\end{tabular}
\end{table}

\end{document}